\documentclass[a4paper]{article}
\usepackage{epsfig,amsmath,amsfonts,amssymb,setspace,multirow,textcomp}
\usepackage[T1]{fontenc}
\usepackage[utf8]{inputenc}
\usepackage[english]{babel}
\makeatletter
\renewcommand{\@oddhead}{\textit{Advances in Astronomy and Space Physics} \hfil}
\renewcommand{\@evenfoot}{\hfil \thepage \hfil}
\renewcommand{\@oddfoot}{\hfil \thepage \hfil}
\makeatother

\renewenvironment{thebibliography}[1]{\begin{oldthebibliography}{#1}\setlength{\parskip}{0ex}\setlength{\itemsep}{0ex}}{\end{oldthebibliography}}

\begin{document}
\fontsize{11}{11}\selectfont 
\title{Evolution of density and velocity profiles of matter \\ in large voids}
\author{\textsl{M.~Tsizh\footnote{tsizh@astro.franko.lviv.ua}, B.~Novosyadlyj}}
\date{\vspace*{-6ex}}
\maketitle
\begin{center} {\small Ivan Franko National University of Lviv, Kyryla i Methodia str., 8, Lviv, 79005, Ukraine}
\end{center}

\begin{abstract}
We analyse the evolution of cosmological perturbations which leads to the formation of large voids in the distribution of galaxies. We assume that perturbations are spherical and all components of the Universe — radiation, matter and dark energy - are continuous media with ideal fluid energy-momentum tensors, which interact only gravitationally. Equations of the evolution of perturbations in the comoving to cosmological background reference frame for every component are obtained from equations of conservation and Einstein's ones and are integrated by modified Euler method. Initial conditions are set at the early stage of evolution in the radiation-dominated epoch, when the scale of perturbation is mush larger than the particle horizon.  Results show how the profiles of density and velocity of matter in spherical voids with different overdensity shells are formed.\\[1ex]
\end{abstract}

\section*{\sc introduction}
\indent Large voids in the spatial distribution of galaxies are elements of large scale structure, study of which can give important information about the hidden components of the Universe - dark matter and dark energy. Usually it is assumed, that dark energy is unperturbed in the voids or, at least, impact of its density perturbations on the peculiar motion and spatial distribution of galaxies is negligibly small. In this paper we investigate the evolution of perturbations of density and velocity of matter together with dark energy ones from the early stage, when the scale of initial perturbation is much larger than particle horizon, up to current epoch. We analyse the influence of dynamical dark energy on such evolution and its dependence on initial conditions. We point attention to the evolution of density and velocity profiles of matter during void formation. For this we have developed the program for integrating the system of equation, obtained for description of evolution of spherical perturbation in 3-
component medium - radiation, matter and dark energy, - from the equations of relativistic hydrodynamics and gravitation \cite{Novosyadlyj16}. The component ``matter'' consists of dark matter (25\% of total density) and typical baryonic matter (5\%) the dynamics of which is well described by the dust-like medium approach in the  large scales. 

\section*{\sc model of spherical void and initial conditions}
\indent We assume that voids in spatial distribution of galaxies are formed as the result of the evolution of cosmological density perturbations with a negative initial amplitude. It is believed that such perturbations are the result of quantum fluctuations of space-time metric in the inflationary epoch. They are randomly distributed in amplitude with normal distribution  and are symmetrical by sign of density perturbation from the average in different regions of space. We consider only scalar mode of perturbations, in which perturbations of density  $\delta_N(t,r)$ and velocity $v_N(t,r)$ in every component $N$ are correlated because of survival of the growing solution  only at the stage when the scale of perturbation was larger than particle horizon. Positive perturbations lead to the formation of galaxies and galaxy clusters and negative ones - to the formation of voids. Formation of structures with positive perturbation is well described by Press-Schechter formalism, theory of Gaussian peaks and halo 
theory of structure formation and their modern modifications based on the numerical N-body simulations. Although the evolution of voids in the distribution of galaxies is much simpler than evolution of galaxy clusters, since it is described by the quasilinear theory, there is no complete theory of voids formation. Here we analyse the development of negative cosmological density perturbations, which form the voids. The mathematical base of their description is the system of 7 differential equations in partial derivatives  for 7 unknown functions of 2 independent variables 
$\delta_{de}(a,r)$, $\delta_m(a,r)$, $\delta_{\rm r}(a,r)$, $v_{de}(a,r)$, $v_m(a,r)$, $v_{\rm r}(a,r)$, $\nu(a,r)$, which were obtained in \cite{Novosyadlyj16} (equations (17)-(22)). Here $\Omega$-s denote the mean densities of the components in the unit of the critical one at the current epoch, $w\equiv p_{de}/\rho_{de}$ is the equation of state parameter of dark energy, $c_s$ is the effective speed of sound of dark energy in its proper frame,  $H(a)\equiv d\ln{a}/dt$ is the Hubble parameter, which defines the rate of the expansion of the Universe and is known function of time for given cosmology and the model of dark energy ($H(a)=H_0\sqrt{\Omega_ra^{-4}+\Omega_ma^{-3}+\Omega_{de}a^{-3(1+w)}}$) and $H_0$ is its today value (Hubble constant). The independent variables are scale factor $a$ and radial comoving coordinate $r$, which define the interval in Friedman-Robertson-Walker 4-space:
\begin{equation}
ds^2=e^{\nu(t,r)}dt^2-a^2(t)e^{-\nu(t,r)}[dr^2+r^2(d\theta^2+\sin^2\theta d\varphi^2)] \label{ds_sph}.
\end{equation}
It is assumed that geometry of 3-space of the Universe (unperturbed cosmological background) is Euclidean. The metric function $\nu(t,r)$ at the late stages, when the scale of perturbation is much smaller than the particle horizon, is the doubled gravitational potential in the Newtonian approximation of eq. (17) in the paper \cite{Novosyadlyj16}. The density and 3-velocity perturbations $\delta_N$  and $v_N$ are defined in coordinates, which are comoving to the  unperturbed cosmological background (see paragraph 2.2 in \cite{Novosyadlyj16}). Thus, the velocity perturbation coincide with definition of peculiar velocity of galaxies (see, for example, \cite{Peebles80}).

To solve the system of equations (17)-(22) from \cite{Novosyadlyj16} the initial conditions must be set. Let us relate the initial amplitude of given perturbation with mean-square one given by power spectrum of cosmological perturbations. For this we define the initial conditions in the early Universe, when $\rho_{\rm r}\gg\rho_{m}\gg\rho_{de}$, and physical size of the perturbation $a\lambda\gg ct$. In that time the perturbations are linear ($\delta,\,v,\,\nu\,\ll\,1$), so without loss of generality the solution can be presented in the form of separated  variables:
\begin{equation}
\hskip-1cm\nu(a,r)=\tilde{\nu}(a)f(r), \quad \delta_{N}(a,r)=\tilde{\delta}_{N}(a)f(r),  
\quad v_{N}(a,r)=\tilde{v}_{N}(a)f'(r), \label{init1} 
\end{equation}
where $f(0)=1$ and $f'(r)\propto r$ near the center $r=0$. Ordinary differential equations for amplitudes $\tilde{\nu}(a)$, $\tilde{\delta}_{N}(a)$, $\tilde{v}_{N}(a)$ are obtained from general system of equations (17)-(22) from \cite{Novosyadlyj16} by their expansion in Taylor series near the center. The analytical solutions of equations for the amplitudes for the radiation-dominated epoch (matter and dark energy can be treated as test components) in  the ``superhorizon'' asymptotic give the simple relation for them: 
\begin{equation}
\tilde{\delta}^{init}_{\rm r}=\frac{4}{3}\tilde{\delta}_{m}^{init}=\frac{4}{3(1+w)}\tilde{\delta}_{de}^{init}=-\tilde{\nu}^{init}=C, \quad \tilde{v}^{init}_{\rm r}=\tilde{v}_{m}^{init}=\tilde{v}_{de}^{init}=\frac{C}{4a_{init}H(a_{init})},\label{init1}
\end{equation}
where $C$ is integration constant, which is defined by initial conditions. We set the value of $C$ in the units of mean-square amplitude of perturbations, which is implied from modern observations. The Planck + HST + WiggleZ + SNLS3 data (see \cite{Sergijenko15} and references therein) tell that amplitude $A_s$ and spectral index $n_s$ of power spectrum of initial perturbations of curvature $\mathcal{P_{R}}(k)=A_s(k/0.05)^{n_s-1}$ are the following \cite{Sergijenko15}: $A_s=2.224\cdot 10^{-9},\,\,n_s=0.963$. Since for perturbations with $ak^{-1}\gg ct$ the power spectrum perturbations of curvature $\mathcal{P_{R}}\equiv<\nu\cdot\nu>$ is constant in time in the matter- and radiation-dominated epochs, in the range of scales $0.01\le k \le0.1$ the initial amplitude which correspondent to mean-square one is: $\sigma_k\equiv\sqrt{A_s}\approx4.7\cdot10^{-5}$. Hereafter we put in our computations $C=-1\cdot10^{-4}\approx2\sigma$ at $a_{init}=10^{-6}$.

\section*{\sc numerical integration}
\indent  For numerical integration of the system of equations (17)-(22) from \cite{Novosyadlyj16} with initial conditions (\ref{init1}) we have created a computer code npdes.f, which implements the modified Euler method taking into account the derivatives from the forthcoming step and improving the results by iterations. This scheme of integration is the most resistant to the numerical spurious oscillations, is the fast and precise enough. For example, the Hamming method of prediction and correction of 4-order of precision with 5 iterations at each step need 3 times more processor time for the same precision of final result. The step of integration  was posed as variable: $da=a/N_a$, where number $N_a$ was picked up so that the numerical precision of the result of integration at $a=1$ was not worse than 0.1\%. In all calculations presented here we took $N_a=3\cdot10^{6}$.   

The numerical derivatives with respect to $r$ in the grid with constant step $dr=R_m/N_r$, where $R_m$ is radius of spatial region of integration, were evaluated with help of 3-rd order polynomial by method of Savitzky-Golay convolution \cite{Savitzky64}: $y'_i=[3(y_{i+1}-y_{i-1})/4-(y_{i+2}-y_{i-2})/12]/dr$. The method was tested by comparing the derivatives of analytical functions of the initial profiles of density and velocity perturbations. The value of step $dr$ was estimated so that the difference between numerical and analytical derivatives do not exceed $\sim10^{-5}$ of their values.

To take into account the Silk damping effect for radiation we have added into equations of evolution of $\delta_{\rm r}$ and $v_{\rm r}$ the terms $\delta_{\rm r}k_D/H/a^2$ and $v_{\rm r}k_D/H/a^2$ accordingly, where the scale of damping $k_D$ was computed by formula (10) from \cite{Hu95}. 

If the values of effective speed of sound in dark energy is $c_s>0.01c$, then the spurious oscillations with growing amplitude appear in this component. Their cause  consist in no perfect scheme of integration by time, the numeric derivatives on spatial coordinates and accumulation of numerical errors. To remove them we used the Savitzky-Golay convolution filter \cite{Savitzky64} with parameters $n_l=12,\,n_r=12,\,m=6$, by which  the space-dependences of derivatives $\dot{\delta}_{de}$ and $\dot{v}_{de}$ were smoothing at each step of integration by $a$. Such smoothing practically does not influence on the final result of integration, which is confirmed by comparison of the results with smoothing and without it for case of the dark energy model with $c_s=0$, for which spurious oscillations do not appear. The maximum difference is less than 4\% for density perturbation and 1\% for velocity perturbation of dark energy in the region of maximum amplitude of velocity perturbation.

The input parameters of the program are: the Hubble parameter $H_0$, the density parameters of all components $\Omega_{\rm r}$, $\Omega_{de}$, $\Omega_m=1-\Omega_{de}-\Omega_{\rm r}$, the equation of state parameter of dark energy $w$, the speed of sound of dark energy $c_s$, the initial amplitude of perturbation $C$, the parameters of profile $f(r)$ of initial perturbation, the parameter of step $N_a$ in $a$, the size of integration region $R_m$ and number of steps of the spatial grid $N_r$.  

The computer code npdes.f has been tested by comparison of the results of the integration by code with
1) known analytical solutions for density and velocity perturbations in conformal-Newtonian frame for radiation- and matter-dominated Universes \cite{Novosyadlyj2007}, 2) results of integration of linear perturbation by CAMB code\footnote{http://camb.info} \cite{camb} and 3) results of integration by  dedmhalo.f code \cite{Novosyadlyj16}, developed on the basis of dverk.f \footnote{http://www.cs.toronto.edu/NA/dverk.f.gz} for perturbation in the central region of the spherical perturbation. In all cases deviations did not exceed a few tenths of a percent, which means, that precision of the integration is better then 1 \%, and hence is high enough for our studies.

\section*{\sc formation of voids in the cosmological models with dark energy}
From our previous studies and studies of other authors we know that the values of density parameter and the equation of state parameter of dark energy are well constrained by current observational data, while the value of effective speed of sound of dark energy is not constrained (see, for example, \cite{Sergijenko15} and citation therein). That is why in this work we analyse the formation of voids in the cosmological models with dark energy with $\Omega_{de}=0.7$, $w=-0.9$ and different values of $c_s\in[0,\,1]$. Other cosmological parameters in computations are fixed too: $\Omega_{\rm r}=4.17\cdot10^{-5}$, $\Omega_m=0.3-\Omega_{\rm r}$, $H_0=70$ km/s$\cdot$Mpc.
\begin{figure}
\includegraphics[width=0.29\textwidth]{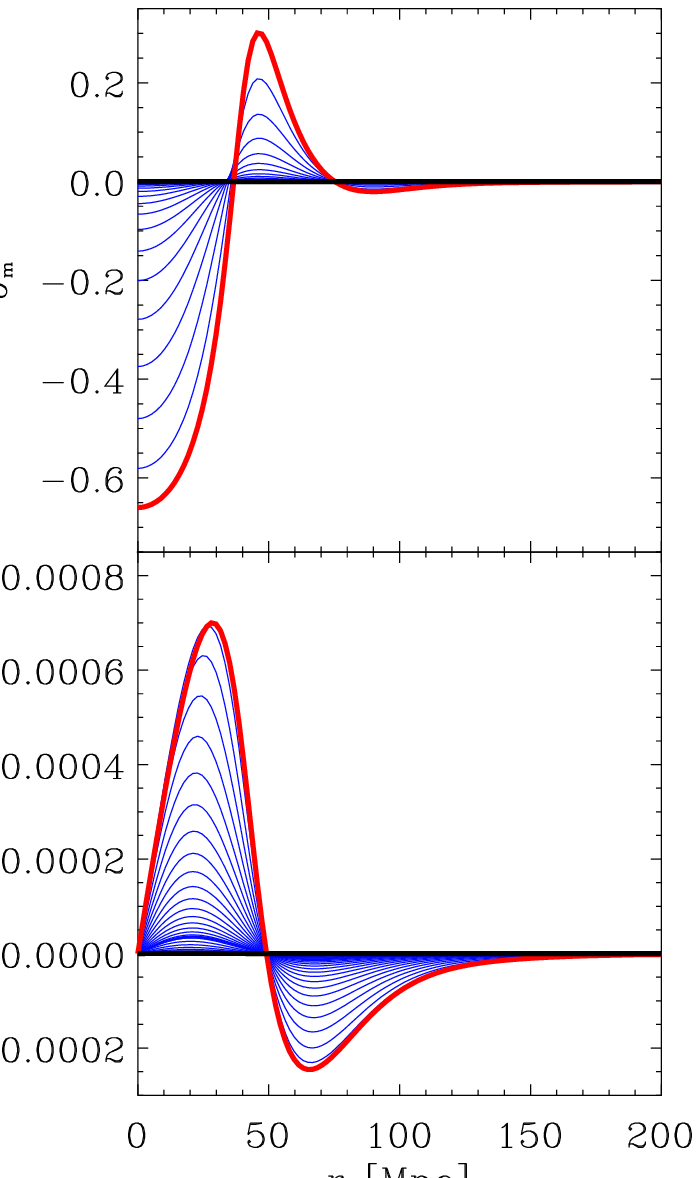} 
\includegraphics[width=0.29\textwidth]{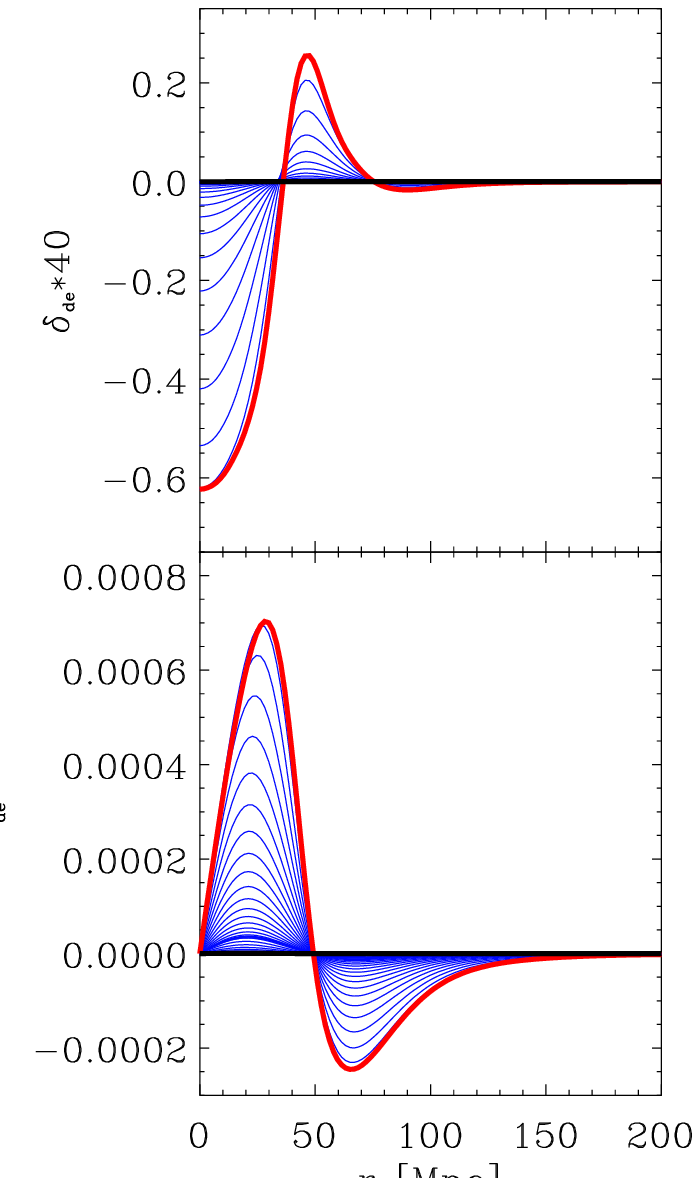}
\includegraphics[width=0.29\textwidth]{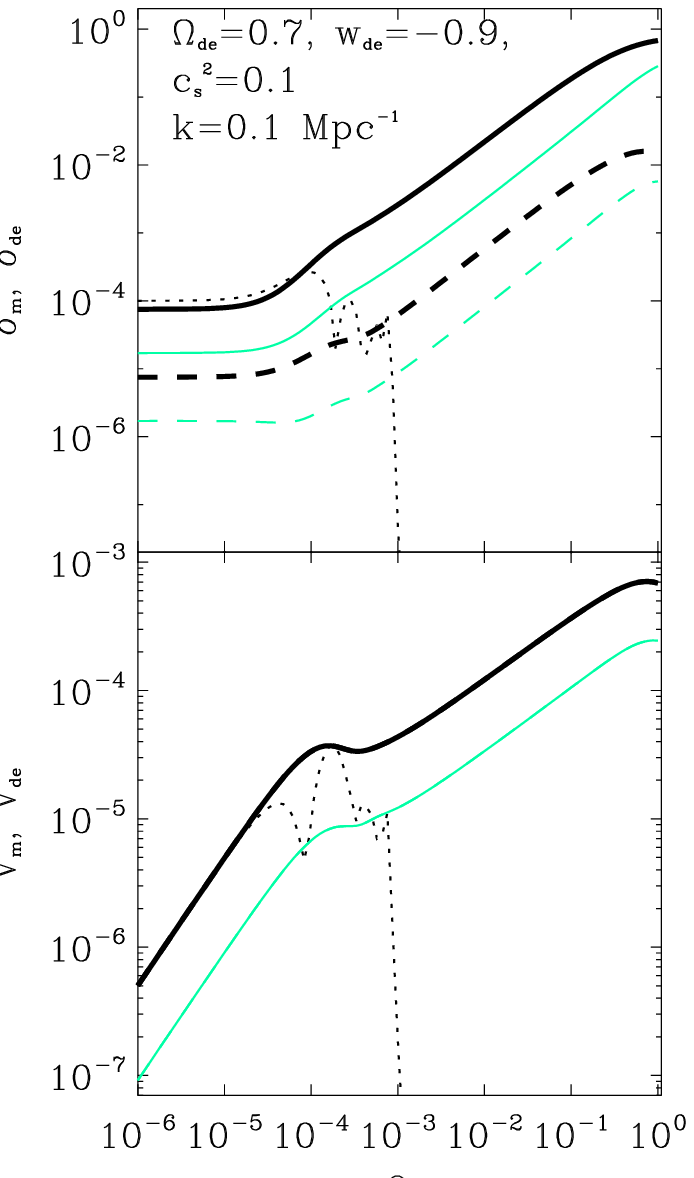}  
\caption{Void formation in dark matter (left column) and dark energy with $c_s^2=0$ (central column). On the right - evolution of absolute values of amplitudes of density (top panel) and velocity (bottom panel) perturbations; solid lines in the top and middle panels - dark matter, dashed lines - dark energy, point lines - radiation; thick black solid and dashed lines - for central point, green ones - for overdensity shell.}
\label{cs00}
\end{figure}
\begin{figure}
\includegraphics[width=0.29\textwidth]{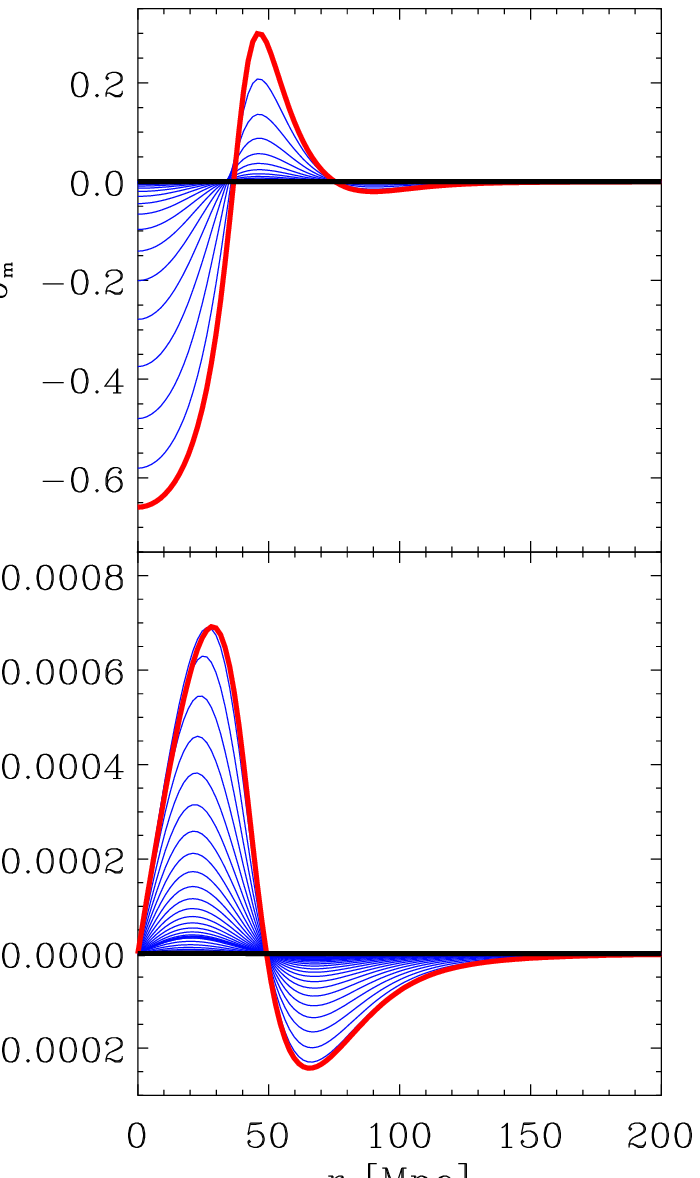} 
\includegraphics[width=0.29\textwidth]{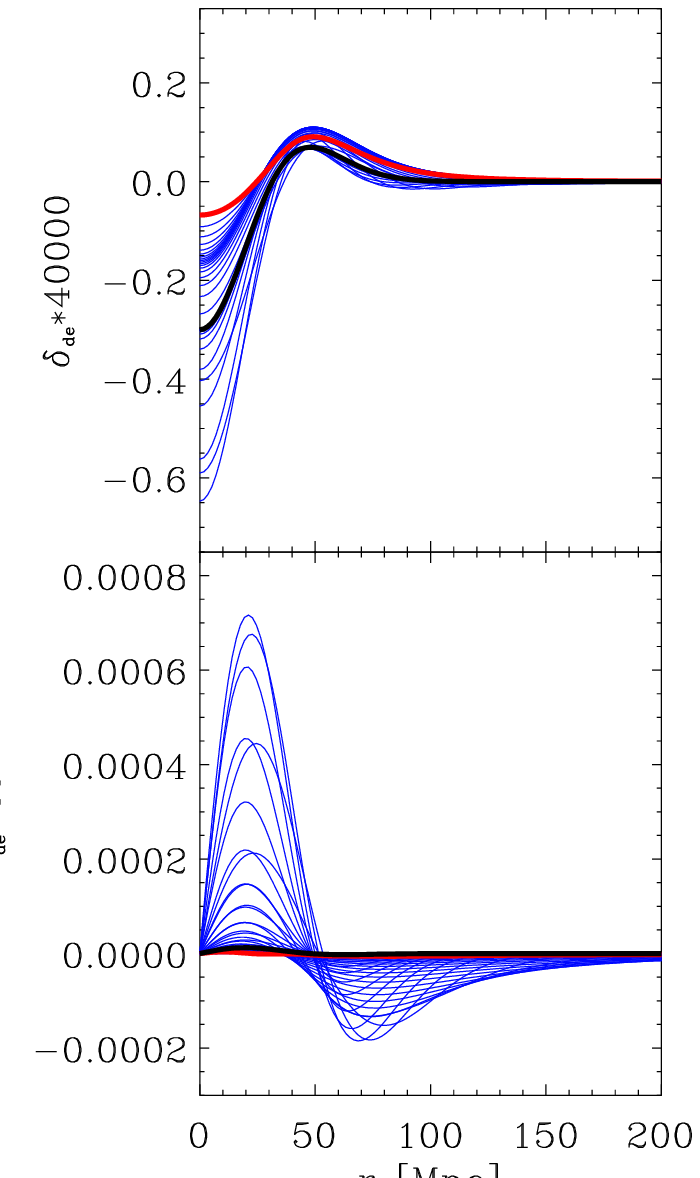}
\includegraphics[width=0.29\textwidth]{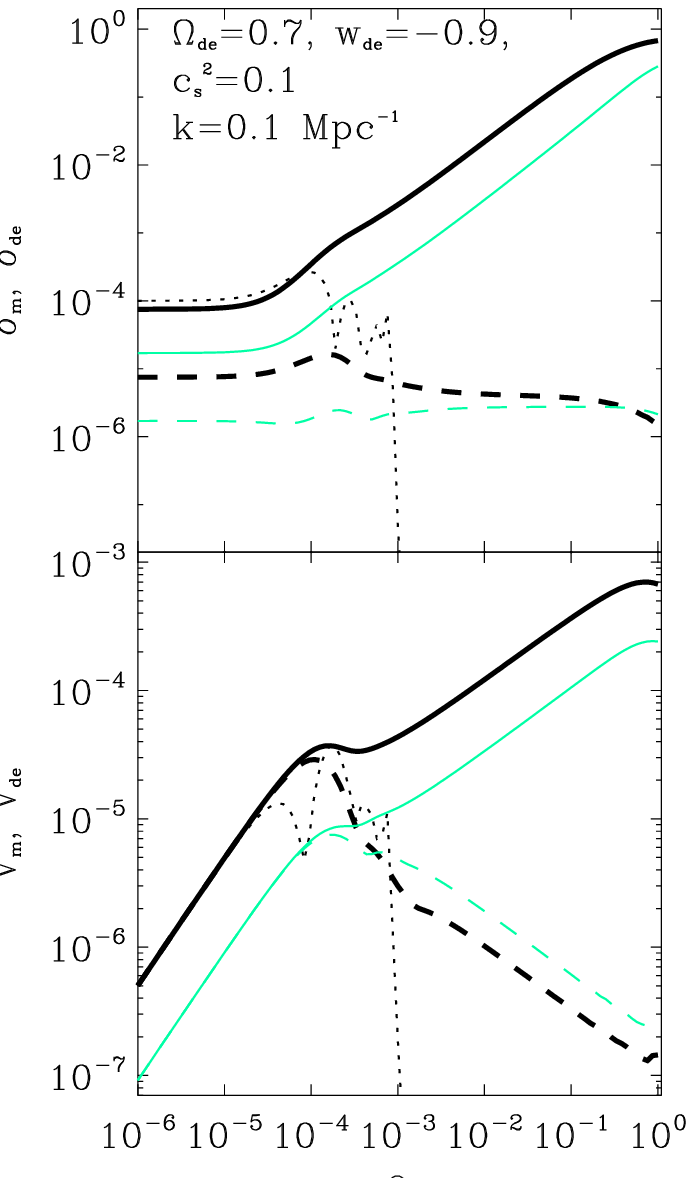}  
\caption{Void formation in dark matter (left column) and dark energy with $c_s^2=0.1$ (central column). On the right - evolution of absolute values of amplitudes of density (top panel), velocity (bottom panel) perturbations in the central point (thick lines) and overdensity shell (thin lines).}
\label{cs01}
\end{figure}

In this work we study the formation of the spherical voids with initial profile $f(r)=(1-\alpha r^2)e^{-\beta r^2}$, where $\alpha$ gives the size of the void  $r_v=1/\sqrt{\alpha}$ and $\beta$ defines the initial amplitude of shell overdensity around the void: $\delta_{e}=-\alpha\beta^{-1}Ce^{-1-\beta/\alpha}$. For comparison of the results of this paper with the results of accompanying one \cite{Tsizh16}, let us set $\alpha=(k/\pi)^2$ and $\beta=3\alpha/4$. This is a proto-void, which is surrounded with overdensity shell with $\delta_{e}\approx\delta(r=0)/8$. For comparison we will also analyse the evolution of the void with shells with smaller amplitudes of overdensity in 2 and 4 times.

In fig. \ref{cs00} we show the formation of the spherical void with $r_v=31.4$ Mpc ($k=0.1$ Mpc$^{-1}$) in the matter and dark energy with $c_s^2=0$: $\delta_{m,de}(a_i,r)$ and $v_{m,de}(a_i,r)$ for $a_i=a_{init}$, ..., $a_{30}=1$. Black lines denote the initial profiles of density and velocity perturbations of both components, red lines denote the final ones. The figure on the right depicts the evolution of absolute values of amplitudes of perturbations in the central point of spherical void (matter - thick solid line, dark energy - thick dashed line) and in the overdensity shell (matter - thin solid line, dark energy - thin dashed line). Velocity perturbation (central panel) are given for the first maximum (thick lines) and first minimum (thin lines). Dotted line denotes the Silk damping for the radiation component.
One can see, that in this dark energy model the perturbations of matter and dark energy grow monotonically after entering the horizon: the black lines are internal, the red lines are external. We also note, that the amplitude of the density perturbation of dark energy is approximately 40 times smaller than the matter one. The values of velocity perturbations of matter and dark energy in this model of dark energy are the same throughout the evolution of the void. They increase monotonically from $a_{init}$ to $a\approx0.56$. It is easy to see that the latter value corresponds to the moment of change from decelerated expansion of the Universe to the accelerated one. The evolution of the absolute values of density and velocity perturbations of matter and dark energy in the overdensity shell is similar to those in the center. 

Similar results of modelling of the void formation in the matter and dark energy with $c_s^2=0.1$ are shown in fig. \ref{cs01}. ``The picture'' of the evolution of the matter density and velocity  perturbations has not changed, while for dark energy it has changed drastically. The final profiles of dark energy perturbations (red lines) are lying on the zero line now. The right figure explains such behaviour of dark energy during the void formation: the velocity perturbation after the entering into horizon decrease quickly, and density perturbation slightly changes during all stages and in the current epoch doesn't differ practically from the background value: $\delta_{de}(1,0)\approx-2\cdot10^{-5}$. The matter density perturbation in the central part of this void at the current epoch is $\delta_m(1,0)\approx-0.7$. We see also that the evolution of the absolute values of density and velocity perturbations of dark energy in the overdensity shell slightly differ from the evolution of ones in the center of the 
void.   

The perturbation of dark energy with larger values of effective speed of sound after entering the particle horizon is smoothed out even faster. Therefore, the ratio of densities of dark energy and matter in the center of the void is
$$\frac{\rho_{de}(1,0)}{\rho_m(1,0)}=\frac{1+\delta_{de}(1,0)}{1+\delta_m(1,0)}\frac{\Omega_{de}}{\Omega_m},$$
and in the case of evolution with considered initial condition this ratio is 3 time larger than on cosmological background. This points to the importance of studying of the voids for establishing  the nature of dark energy.  

\begin{figure}
\includegraphics[width=0.29\textwidth]{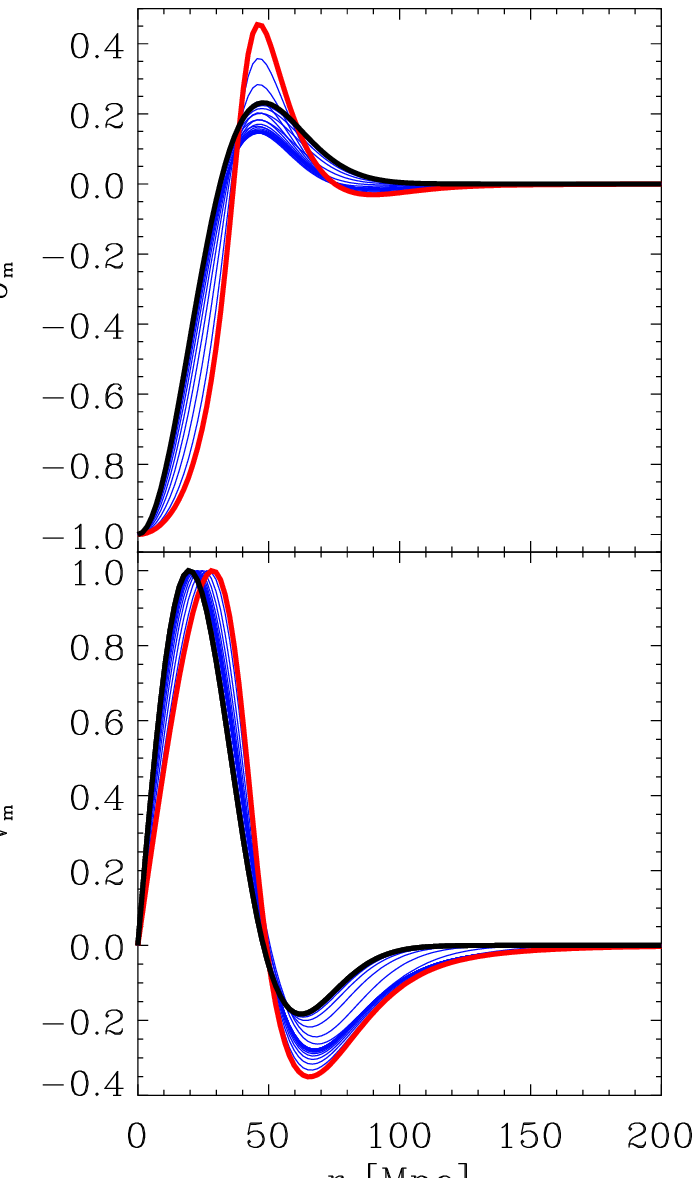} 
\includegraphics[width=0.29\textwidth]{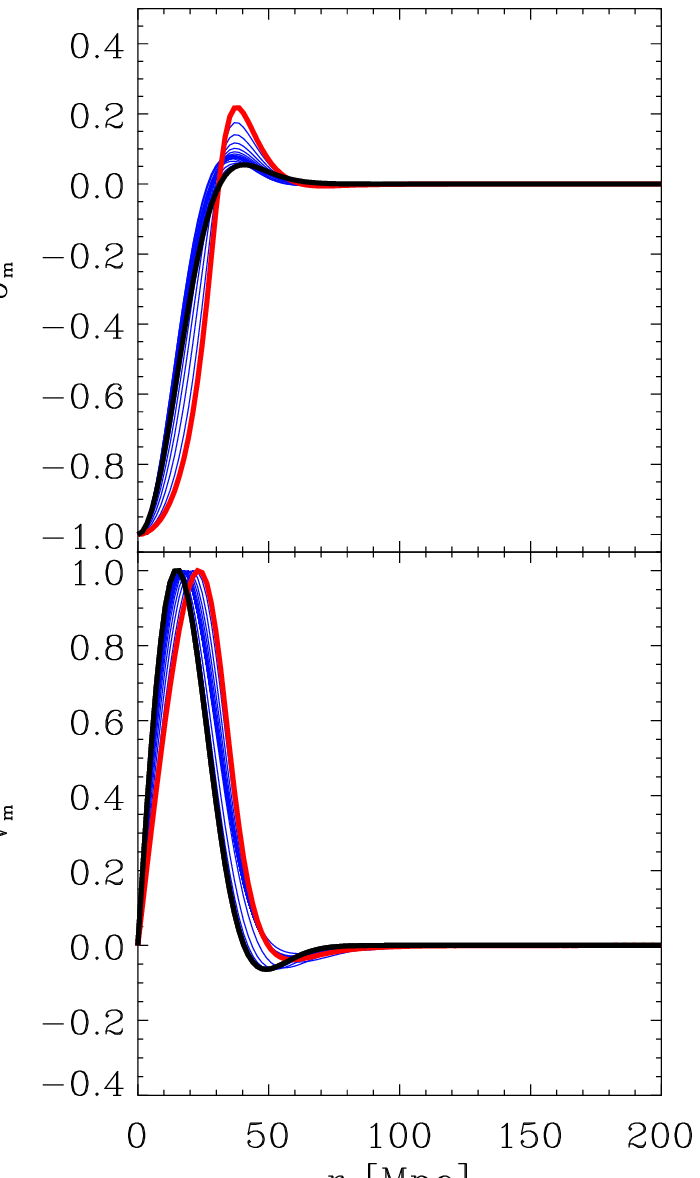}
\includegraphics[width=0.29\textwidth]{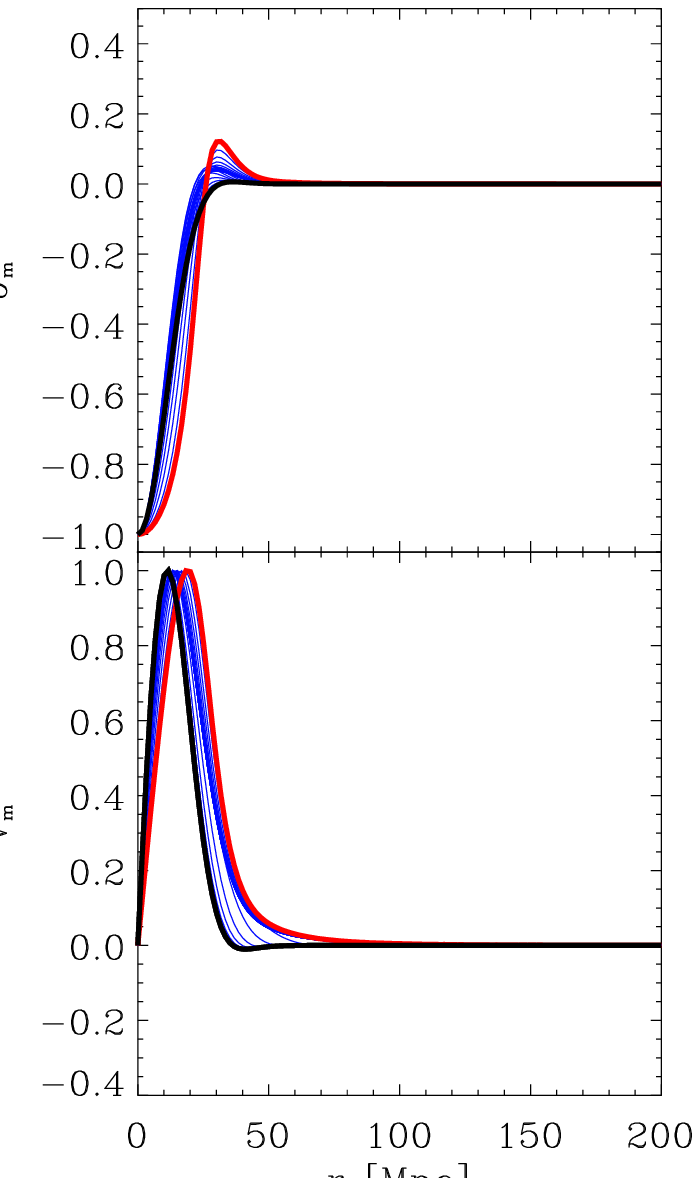}  
\caption{Evolution of profiles of matter density and velocity perturbations with initial parameters $\alpha=(31.4)^{-2}$ Mpc$^{-2}$ and $\beta=3\alpha/4$ (left), $\beta=3\alpha/2$ (central), $\beta=3\alpha$ (right). Thick black line is initial profile, thick red one is final profile.}
\label{profiles}
\end{figure}

Study of the evolution of spatial profiles of matter density and velocity perturbations is important for understanding of the formation of voids. They can be obtained by normalization of every curve in the left columns in fig. \ref{cs00}-\ref{cs01} by its amplitude. The result is given in figs. \ref{profiles} for initial profiles with three different values of parameter $\beta$. 
They show, that in our model the perturbation with initial density profile $\delta_{init}(r)=-1\cdot10^{-4}[1-(r/r_v)^2]e^{-\beta r^2}$ with $r_v=31.4$ Mpc and $\beta=3r_v^{-2}/4$ (left column) leads to the formation of void with radius in comoving coordinates $\approx38$ Mpc with amplitude of density perturbation in the center $\delta_m(1,0)\approx-0.68$ and the overdensity shell around it ($\delta_m>0$) with thickness $\approx36$ Mpc and amplitude of density perturbation $\delta_e\approx0.33$. In the case of $\beta=3r_v^{-2}/2$ (central column) the radius of the central void is $\approx31$ Mpc, the amplitude of density perturbation in the center is $\delta_m(1,0)\approx-0.69$, the shell of overdensity has thickness $\approx30$ Mpc and the amplitude of density perturbation $\delta_e\approx0.16$. In the case of $\beta=3r_v^{-2}$ (right column) the void has the following parameters: the radius $\approx25$ Mpc, the  amplitude of density perturbation in the center $\delta_m(1,0)\approx-0.73$, the maximum of 
overdensity in the shell  $\delta_e\approx0.09$ is at distance $\approx31$ Mpc from the center, external bound of shell, where sign of perturbation changes from ''+`` to ''-`` is absent.  
Important is the dependence of profiles of peculiar velocity of matter in the void and around it on the model parameters and initial perturbation. From fig. \ref{cs00}-\ref{profiles} one can see, that the first positive peak of peculiar velocity (from the center) is at the edge of the void and the second negative (velocity towards the center) is at the edge of the overdensity shell. The values of velocities in the case of $\beta=3r_v^{-2}/4$ are $v_{m-v}(30\,\mathrm{Mpc})\approx200$ km/s, $v_{m-e}(65\,\mathrm{Mpc})\approx-70$ km/s,
in the case of $\beta=3r_v^{-2}/2$: $v_{m-v}(25\,\mathrm{Mpc})\approx170$ km/s, $v_{m-e}(60\,\mathrm{Mpc})\approx-66$ km/s, 
and in the case of $\beta=3r_v^{-2}$: $v_{m-v}(20\,\mathrm{Mpc})\approx150$ km/s and the second negative peak is absent.  

Note, that final value of the amplitude of the perturbation in the shell is the greater, the greater is its initial value (smaller value of $\beta$), for the same value of initial amplitude in the center. One can see also, that overdensity shell appears in the process of evolution of void even if its amplitude was very small in the initial profile (figure on the right), or absence at all ($\alpha=0$, Gaussian initial profile). The evolution of matter density and velocity profiles points that for interpretation of the observational data on the distribution of void galaxies in the phase space the non-linear theory should be used (see also table in \cite{Tsizh16}).     

\section*{\sc conclusion}
The large voids in the spatial distribution of galaxies are formed from the negative cosmological density perturbations of matter. The amplitude of the density perturbation in the central part of the void at the current epoch is defined by the depth of dip of Gaussian field of the initial matter density perturbations, the parameters of the cosmological model and parameters of the initial profile. For example, in the cosmological model with quintessential dark energy the initial negative density perturbation with profiles similar to the Gaussian ones lead to the formation of the voids with the overdensity shells. In such voids with $r_v\approx30$ Mpc and $\delta_m(1,0)\approx-0.7$ the maximal values of the peculiar velocity of galaxies are $\sim150-200$ km/s (movement from the center in the comoving coordinates) and are reached near the boundary. In the shells such velocity is directed to the center, however its value does not exceed $\sim70$ km/s. With increasing the parameter of initial profile $\beta$ for 
the same $r_v$ the amplitudes of the density and velocity perturbations in the shells decrease.

The density and velocity perturbations of the dark energy evolve similarly to the perturbations of matter at the stage when their scales are much larger than the particle horizon. After they enter the particle horizon their evolution depends on the value of the effective speed of sound $c_s$. If $c_s=0$, then similarity is conserved with the difference that the amplitude of density perturbation of dark energy is smaller in factor $1+w$. At the later epoch, when the dark energy density dominates, this difference increased yet in $\approx4-5$ times more. If $0<c_s\le1$, then the amplitude of velocity perturbation of dark energy after entering the horizon decreases rapidly, the amplitude of the density perturbation doesn't increase or even decreases too. Therefore, in the voids the density of quintessential dark energy is approximately the same as in cosmological background. The ratio of the densities of dark energy and matter is in  $1/(1+\delta_m)$ larger than in the cosmological background. The more hollow 
void is the larger this ratio is. That is why the large voids are important elements of large-scale structure of the Universe for testing models of dark energy and gravity modifications.  


\end{document}